\documentclass[onecolumn,natbib,lscape]{aa}
\usepackage{natbib}
\usepackage{graphicx}
\usepackage{graphics}
\usepackage{longtable}
\usepackage{lscape}
\usepackage{amssymb}
\usepackage{txfonts}
\def\arcsec{\ifmmode^{\prime\prime}\;\else$^{\prime\prime}\;$\fi}
\def\arcmin{\hbox{$^\prime$}}
\def\deg{\hbox{$^\circ$}}
\def\magcir{\ \raise -2.truept\hbox{\rlap{\hbox{$\sim$}}\raise5.truept
\hbox{$>$}\ }}                        
\def\mincir{\ \raise -2.truept\hbox{\rlap{\hbox{$\sim$}}\raise5.truept
\hbox{$<$}\ }}                        

\begin{document}
         \title{Unobscured QSO\,2: A New Class of Objects?\footnote{Based 
             on observations taken at the Italian Telescopio Nazionale 
             Galileo, TNG and on observations obtained with XMM-Newton, 
             an ESA science mission with instruments and contributions 
             directly funded by ESA member states and NASA.}}
        \author{A. Wolter,
              \inst{1} 
	      I. M. Gioia,
              \inst{2}
	      J. P. Henry, 
              \inst{3} and
	      C. R. Mullis
              \inst{4}
}
\offprints{A. Wolter}
    \institute{INAF - Osservatorio Astronomico di Brera, Via Brera 28, I-20121
	   Milano, Italy \\
         \email{anna@brera.mi.astro.it}
                        \and
   INAF-CNR - Istituto di Radioastronomia, Via Gobetti 101, 
            I-40129 Bologna, Italy \\
         \email{gioia@ira.cnr.it}  
                        \and
 Institute for Astronomy, University of Hawai$'$i, 2680 Woodlawn Drive, Honolulu, HI 96822, USA \\
          \email{henry@ifa.hawaii.edu}		
                       \and
 Department of Astronomy, University of Michigan, 918 Dennison Building, Ann Arbor, MI 48109, USA \\
 		\email{cmullis@umich.edu}
}

\date{Received ......... 2005; accepted ......... 2005}

   \abstract{
We present in this paper optical and X-ray follow up observations
for three X-ray selected objects extracted from the  {\em ROSAT}
North Ecliptic Pole survey which is a flux-limited, completely 
identified survey. All three objects have X-ray luminosities in the 
10$^{44}$  erg s$^{-1}$ regime and show narrow  emission lines in their 
optical discovery spectra, typical of  QSO\,2 type objects. Spectroscopic
data for the three QSO\,2 candidates, obtained with the Telescopio 
Nazionale Galileo, confirm the widths of the H$\alpha$ or H$\beta$ 
emission lines are less than 750 km s$^{-1}$. On the other hand 
XMM-{\em Newton} data do not show any sign of obscuration as expected 
for this class  of objects. The X-ray spectra of the three objects
are all well fit by a single power law model with $\Gamma\sim$1.7 
with low energy absorption fixed to the Galactic value along the line 
of sight to each object. Most observational evidence supports 
the scenario where optical and X-ray obscurations are linked, 
contrary to our findings. We discuss the unanticipated results of 
these observations, and compute the  space density in soft  X-ray surveys
of this possibly new class of objects. Their spatial density in the 
{\em ROSAT} NEP survey is 2.8$^{+2.7}_{-1.5} \times10^{-8}$ h$^{3}$ 
Mpc$^{-3}$  in a $\Lambda$CDM model with $h$$=$0.7.  Unobscured QSO\,2 
candidates could go unrecognized  in current  X-ray surveys where the 
low hydrogen column density is inferred by a hardness ratio rather than 
a more precise  X-ray spectrum measurement. 

\keywords{galaxies: active; galaxies: quasars: general - X-rays;  
X-rays: individual: RX\,J1715.4+6239, RX\,J1724.9+6636, RX\,J1737.0+6601}
}
\authorrunning{Wolter et al.}
\titlerunning{Unobscured QSO\,2}

   \maketitle
%

\section{Introduction}
The phenomenology of Active Galactic Nuclei (AGN) is extremely varied with 
luminosities ranging over many orders of magnitude. AGN have been classified 
over the years into several subclasses to explain the variety of features 
observed like the presence or lack of broad optical  emission lines, 
the presence and morphology of radio emission. Historically, low 
luminosity AGN are called Seyferts (Sy), and high luminosity AGN are 
called quasars (QSO).  Objects are further classified  as Type 1 or Type 
2 according to the presence of broad or narrow optical emission lines.
A recently defined class of objects are the Type 2 QSO, or QSO\,2, which 
are heavily obscured, powerful hard X-ray emission objects. They are 
expected to be the luminous counterparts of the local Sy\,2 galaxies with 
narrow optical emission lines.

\medskip\noindent
Absorbed AGN are a major fraction in the total emission budget of the diffuse 
X-ray Background \citep[XRB;][]{sch98}, as was first predicted by \cite{sw89}.
The model was later refined by  \cite{mgf94},  \cite{com95}, \cite{grs99}, 
among others. In more recent years the XRB 
was directly resolved with Chandra by \cite{mush00},  \cite{bran01},
\cite{giac01}, \cite{ros02},  and with XMM-Newton by \cite{has01}.
The original requirement of the Setti \& Woltjer model, that the 
number of absorbed sources is approximately equal to that of unabsorbed ones,
still holds.  QSO\,2 are deemed necessary to explain the hard source counts 
at fluxes of the order of 10$^{-13}$ erg cm $^{-2}$ s$^{-1}$ in the
ASCA 2$-$10 keV  band where  they should contribute  at least 30\% of the 
total emission \citep{gsh01}.
The relative densities of absorbed and unabsorbed objects in the
model can be compared  to the relative densities found in the optical band. 
The local ratio  between Sy\,2 and Sy\,1 in the optical band ranges from 2 
\citep{hb92} to 4 \citep{mr95}. Since there is no reason to expect a 
different ratio for high luminosity objects, we should  expect a factor of 
2-4 more QSO\,2 than QSO\,1. What is observed instead is a  dearth of 
QSO\,2 \citep[see among others:][]{aki02,rdc03,bar98,fio99}.

\medskip\noindent
Since QSO\,2 are a key ingredient in models of the diffuse  XRB 
\citep{com00,gsh01}, they should be more easily detected in X-ray 
selected  samples because they have high X-ray luminosities. 
No X-ray survey found a sizeable number of such objects  
\citep[e.g. there are no QSO\,2 in the complete ASCA Large Sky 
Survey;][]{aki00}, and only a few candidates have been found using ASCA 
\citep{oht96,aki02,rdc03}, ~{\em ROSAT} \citep{alm95,bar98,geor99},
and {\it Beppo}-SAX \citep{fio99}.
Detection of QSO\,2 is very controversial. Several QSO\,2 candidates  
have been proposed in the literature, but they have often been classified 
later with other known types of objects. This is the case of AXJ0341.4$-$4453,
proposed as a Type\,2 AGN by \cite{boy98}, and later identified as a Narrow 
Line Sy\,1  \citep{hal99} on the basis of a broad component in the 
H$\beta$ emission line (not observed in the discovery spectrum) plus other 
spectral properties \citep[see for the classification of Narrow Line 
Sy\,1:][]{ost85,good89}. 
{\em CHANDRA}  and XMM-{\em Newton} data  have included new candidates 
\citep{ster02,nor02,daw01,mai02,fio03,cac04} 
but the number of confirmed QSO\,2 is 
still small. Their absence in X-ray selected samples could be explained if 
``all'' sources are perfectly Compton-thick 
(N$_{\rm H} \magcir 10^{24}$ cm$^{-2}$), implying a complete absorption 
of the X-ray  radiation. Alternatively,  the luminosity of the nucleus could 
be so high that  the obscuring material is  effectively removed from the  
circumnuclear regions, leaving a totally  unblocked view of the Broad Line  
region (BLR). Both interpretations seem unlikely. It has been
suggested \citep{ris03} that some soft X-ray weak quasars with red
optical counterparts might be optically broad-lined objects with an
intrinsically weak and flat X-ray spectrum. 
This model seems to be supported by the fact that there is no 
one-to-one correspondence between optical and X-ray absorption. A number of 
galaxies, optically classified as Sy\,1, shows high intrinsic hydrogen 
columns  
or at least soft energies spectral curvatures that might be attributed
to ionized absorbers \citep{np94,fio01}. 
Also a number of objects, 
optically classified as Sy\,2, shows absorptions smaller than 10$^{22}$ 
cm$^{-2}$ even if they might still be reconciled with the unifiying scenario 
\citep[see e.g.][]{gua01,pap01,pb02}.  

\medskip\noindent
The comparisons of optical and X-ray  absorption has not yet been done 
with a large statistical sample. This 
will be possible in the near future for instance with the new samples 
identified by the XMM Survey Science Center \citep{rdc04} or from surveys 
with {\em CHANDRA} \citep{sil05}. First results seem to confirm 
that  the optical and X-ray absorption are linked \citep{cac04}.
Therefore objects optically appearing as Type\,2 should  present a high 
degree of X-ray absorption.
We have selected from the {\em ROSAT} North Ecliptic Pole survey 
\citep[NEP;][]{gio03} three QSO\,2 candidates on the basis of high X-ray 
luminosity and absence of broad optical emission lines. We present in this 
paper new optical and X-ray observations of the three NEP QSO\,2 candidates 
present in the survey catalog and discuss the results obtained.  We address 
the  issue of the density of QSO\,2 and their spectral properties in a 
completely identified, X-ray-flux  limited sample. 
Throughout the paper quoted  uncertainties are  90\% confidence levels for 
one  interesting parameter.  We assume the currently-favored  cosmological 
concordance model  ($h, \Omega_M, \Omega_\Lambda$)=(0.7, 0.3, 0.7).
\begin{center}
\begin{table*}
\caption{{\em ROSAT} NEP Properties of the three QSO\,2 Candidates}
\label{tab1}
\begin{center}
\begin{tabular}[h]{l l c  c c c c c}
\hline
\hline
Source Name & NEP id \#  & PSPC F$_{0.5-2.0 keV} \pm \Delta$F & z (UH or MMT)  & z (TNG) & N$_{\rm H}$ & L$_{0.5-2.0 keV}$  & M$_{\rm B}$ \\
       &  & 10$^{-13}$ erg cm $^{-2}$ s$^{-1}$      &      &  & cm$^{-2}$  & 10$^{44}$ erg s$^{-1}$ & (from USNO B mag)\\
\hline
\\
RX\,J1715.4+6239 & NEP\,1239 & 1.39$\pm$0.30 & 0.8500 & 0.8500 & 2.62$\times 10^{20}$ &4.88  & -23.5\\
          & &           &       &            &            \\
          & &           &       &            &            \\
RX\,J1724.9+6636 & NEP\,1640 & 1.03$\pm$0.25 & 0.6792 & 0.6758 & 3.84$\times 10^{20}$ &2.08 & -23.2\\
          & &           &       &            &             \\
          & &           &       &            &             \\
RX\,J1737.0+6601 & NEP\,2131 & 1.36$\pm$0.20 & 0.3580 & 0.3585 & 3.76$\times 10^{20}$ &0.60  & -24.1\\
         &  &           &       &            &             \\
         &  &           &       &            &             \\
\hline 
\end{tabular}
\end{center}
\end{table*}
\end{center}

\section{The NEP sample of QSO\,2 candidates}

The {\em ROSAT}  data around the North Ecliptic Pole of the {\em ROSAT}
All Sky Survey ({\em RASS\/}) have been used to construct a contiguous 
area survey consisting  of a homogeneous X-ray flux-limited sample of 
445 X-ray   sources \citep{mul01,hen01,vog01,gio03}. 
The region around the NEP possesses the deepest 
exposure and consequently the greatest  sensitivity of the entire 
{\em RASS\/} \citep{vog99}.  Hence,  the 9\deg$\times$ 9\deg 
~survey region centered at $\alpha_{2000}=$ 18$^{h} 00^{m}$,
~$\delta_{2000}=+$66\deg 33\arcmin ~covers the deepest, wide-angle  
contiguous region ever observed in X-rays.
This unique combination of depth plus wide, contiguous solid angle provides
the capabilities of detecting both high-redshift objects and large scale
structure. A comprehensive program of optical follow-up observations to 
determine the nature of each of the X-ray sources in the NEP sample led
to the identification of all but two NEP sources \citep{gio03}. 
There is evidence that at least one of the two unidentified sources is a 
blend and that both may be spurious. 

\medskip\noindent
The main constituents of the catalog are AGN (49\%), either Type\,1 or 
Type\,2 according to  the broadness of their permitted emission lines 
with the  cut-off at FWHM$=$2000 km s$^{-1}$ for classification purposes. 
There are 198 AGN\,1 and 21 AGN\,2, where under the term AGN we include 
both Seyfert and QSO. We further divide QSO and Seyfert galaxies
according to their X-ray luminosity and put the dividing line  at 
L$_{X} \sim 10^{44}$ erg s$^{-1}$ in the $0.5-2.0$ keV energy band. Only 
three of a total of 219  AGN are QSO\,2 candidates. These objects have narrow 
permitted emission lines  (MgII 2798 \AA\ and H$\gamma$ with FWHM$\la$1000 
km s$^{-1}$) and show high-ionization emission lines like [NeIII] 3869 \AA\ 
(see two of the spectra in Fig.~\ref{fig1} (top) and Fig.~\ref{fig2}). 
The 99.6\% identification rate implies that QSO\,2 are  a relatively small  
fraction of the total  AGN population, at least in the  soft X-ray 
$0.5-2.0$ keV energy band. The properties of the three candidates are given 
in  Table~\ref{tab1}. The explanation of each column is given below:
\begin{itemize}

\item Columns (1) and (2): Object  {\em ROSAT} name and internal NEP survey 
            identification number.
\item Column (3): Unabsorbed PSPC flux and its error in the standard 
             {\em ROSAT}  energy band (0.5-2.0 keV)
             derived from the count rate 
            assuming the source has a power law spectrum (photon index $=$ 2),
            with absorption fixed at the galactic value for the source 
            position.
\item Columns (4) and (5): Redshift as measured at various telescopes, 
              UH stands for the University of Hawai$'$i  2.2m telescope, MMT 
              for the Multiple Mirror 6.5m  Telescope and TNG for the 3.6m 
              Telescopio Nazionale Galileo. 
\item Column (6) Column density of Galactic hydrogen  along the line 
              of sigth to each object from \cite{elv94} with supplements from 
             \cite{star92}.
\item Column (7): Rest-frame, K-corrected luminosity in the 0.5--2.0 keV 
             energy band. Note that K-correction for a power law  spectrum
             is $(1+z)^{\Gamma-2}$ and is thus unity for $\Gamma=2$.
\item Column (8): Absolute Blue Magnitude from brightest B magnitude in USNO-B1 ({\em http://www.nofs.navy.mil/data/fchpix/ }). 
\end{itemize}
The luminosity derived for these objects, L$_{X} \magcir 10^{44}$ erg 
s$^{-1}$, as well as their optical magnitude are in the QSO class range. 
Variations of B and R magnitudes from the USNO catalog at different epochs,
with differences ranging from 0.1 up to 1.6 mag, show the variability
nature of these objects consistent with the AGN classification.

\tabcolsep 0.7cm
\begin{center}
\begin{table*}
\caption{Rest-frame Optical Emission Line Widths}
\label{tab2}
\begin{center}
\begin{tabular}[h]{l r r r r r}
\hline
\hline
Line id          & NEP\,1239 & NEP\,1640     & NEP\,2131 \\
                    &km s$^{-1}$& km s$^{-1} $  & km s$^{-1}$ \\
\hline
\\
$[$NeV$]$ 		    &		&		& 819  \\
$[$OII$]$ 		    &		&		& 801  \\
$[$NIII$]$		    &		&		& 460  \\
H$\gamma$	    &		&		& 633  \\
H$\beta$            & 453	& 270           &      \\
$[$OIII$]\lambda$4958 & 458	& 480           & 741  \\
$[$OIII$]\lambda$5007 & 457	& 623           & 707  \\
$[$NeI$]$               &           & 608	        &      \\
H$\alpha$           &      	&               & 726  \\
\\
\hline  
\end{tabular}
\end{center}
\end{table*}
\end{center}
\section{Observations and Data Analysis}

In this section we present the discovery optical spectra for two of the three
QSO\,2 candidates obtained at the University  of Hawai$'$i (UH) 2.2m telescope
and the new optical spectroscopy performed at the Italian 3.6m Telescopio 
Nazionale Galileo (TNG), plus the X-ray follow-up observations  acquired with
XMM-{\em Newton}.  We first  describe  the optical spectroscopy,
and then present the X-ray data.

\subsection{Optical Spectroscopic Data}

\begin{figure}
   \centering
   \includegraphics[bb=0 100 574 654, width=14cm]{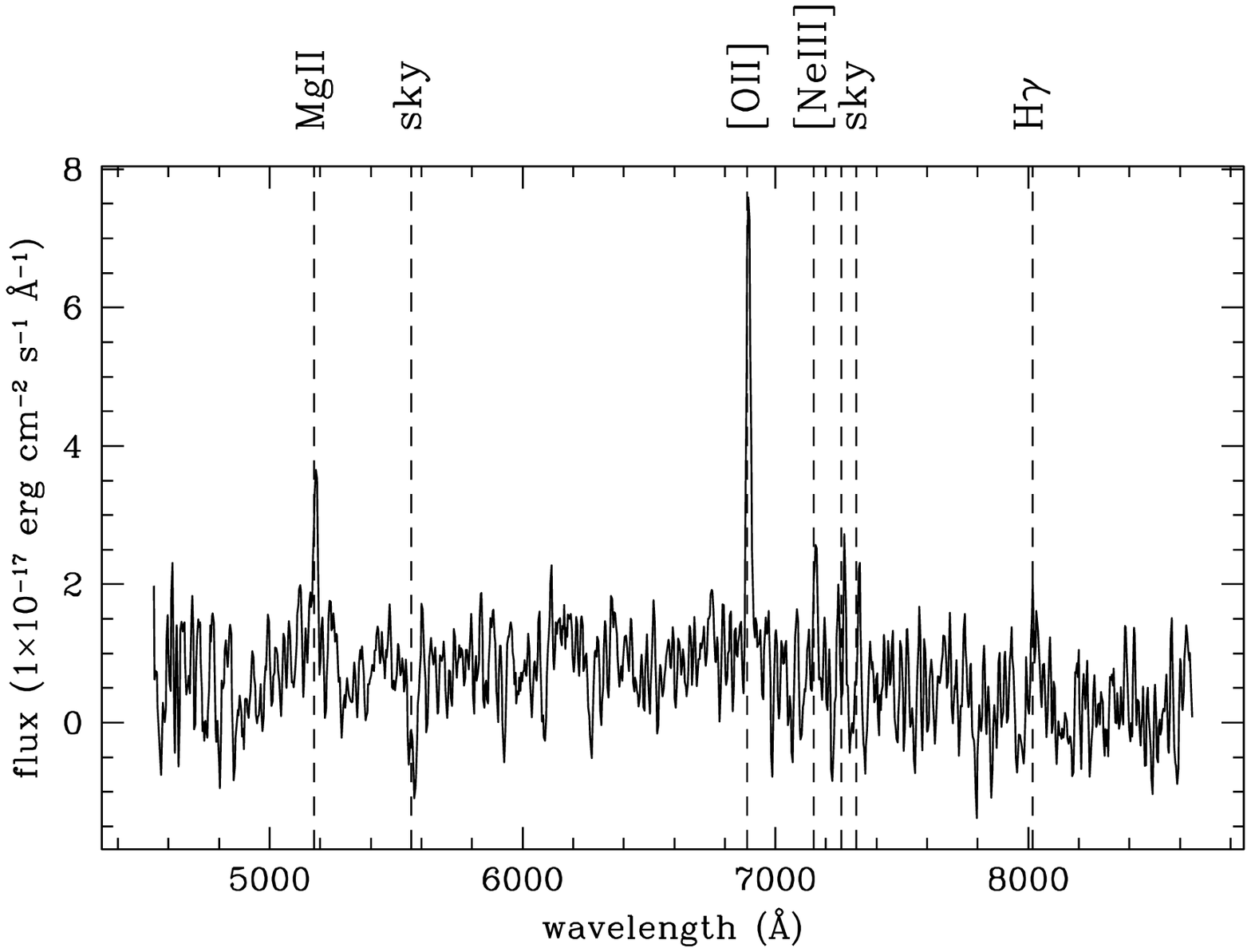}
   \includegraphics[bb=0 100 574 584, width=14cm]{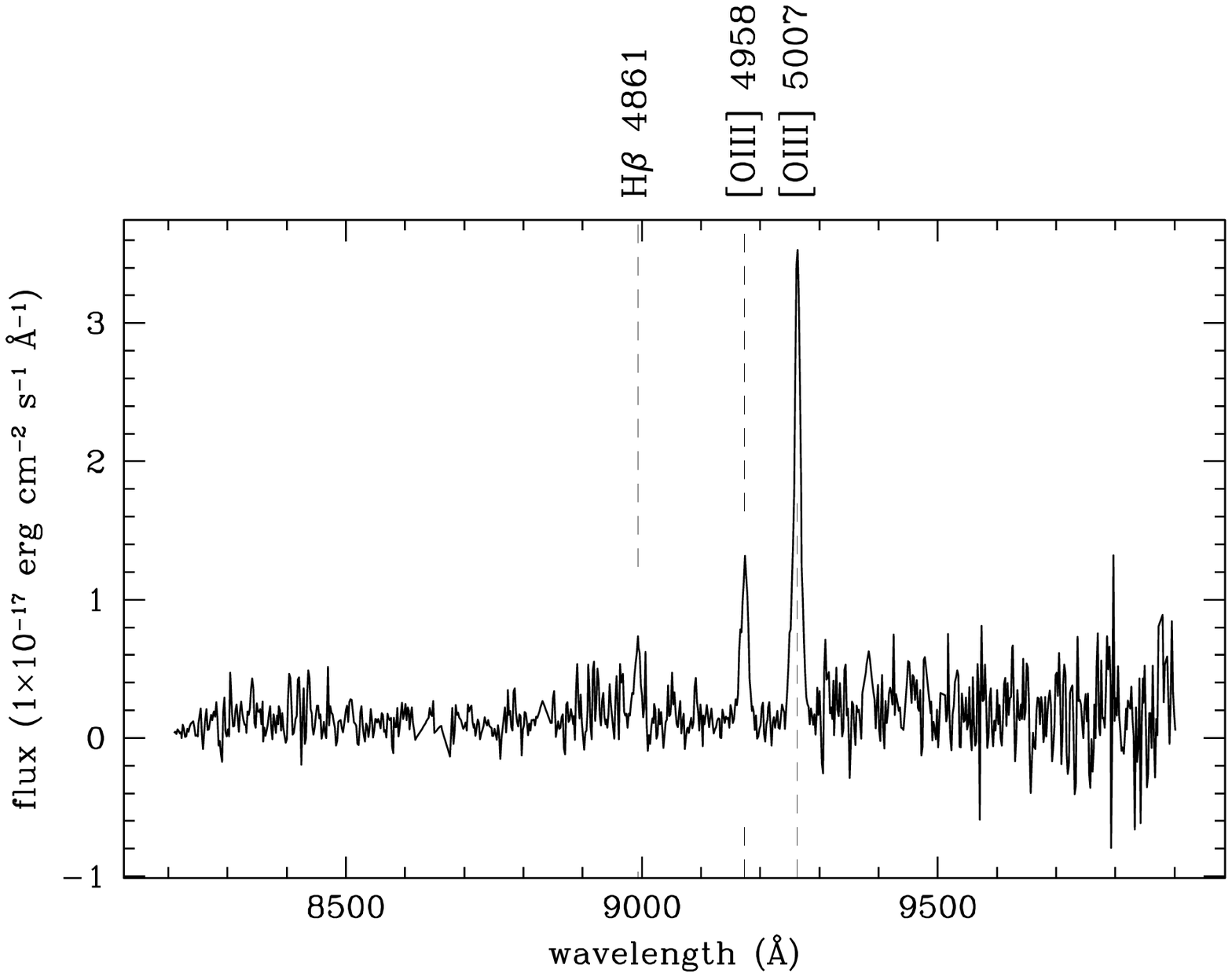}
   \vspace{-3cm}
   \caption{NEP\,1239: (top) longslit spectrum of RXJ\,1715.4$+$6239, QSO\,2
     candidate at z$=$0.85, obtained with the University of Hawai$'$i 2.2m 
     telescope; (bottom) longslit spectrum of the same object at redder
     wavelengths taken at the TNG 3.6m telescope. The dashed lines indicate 
     the positions of the emission lines at the object redshift. Wavelengths 
     of atmospheric absorption (or bad sky subtracion) are also indicated.}
\label{fig1}
\end{figure}

\begin{figure}
   \centering
  \includegraphics[bb=0 0 574 630, width=10cm]{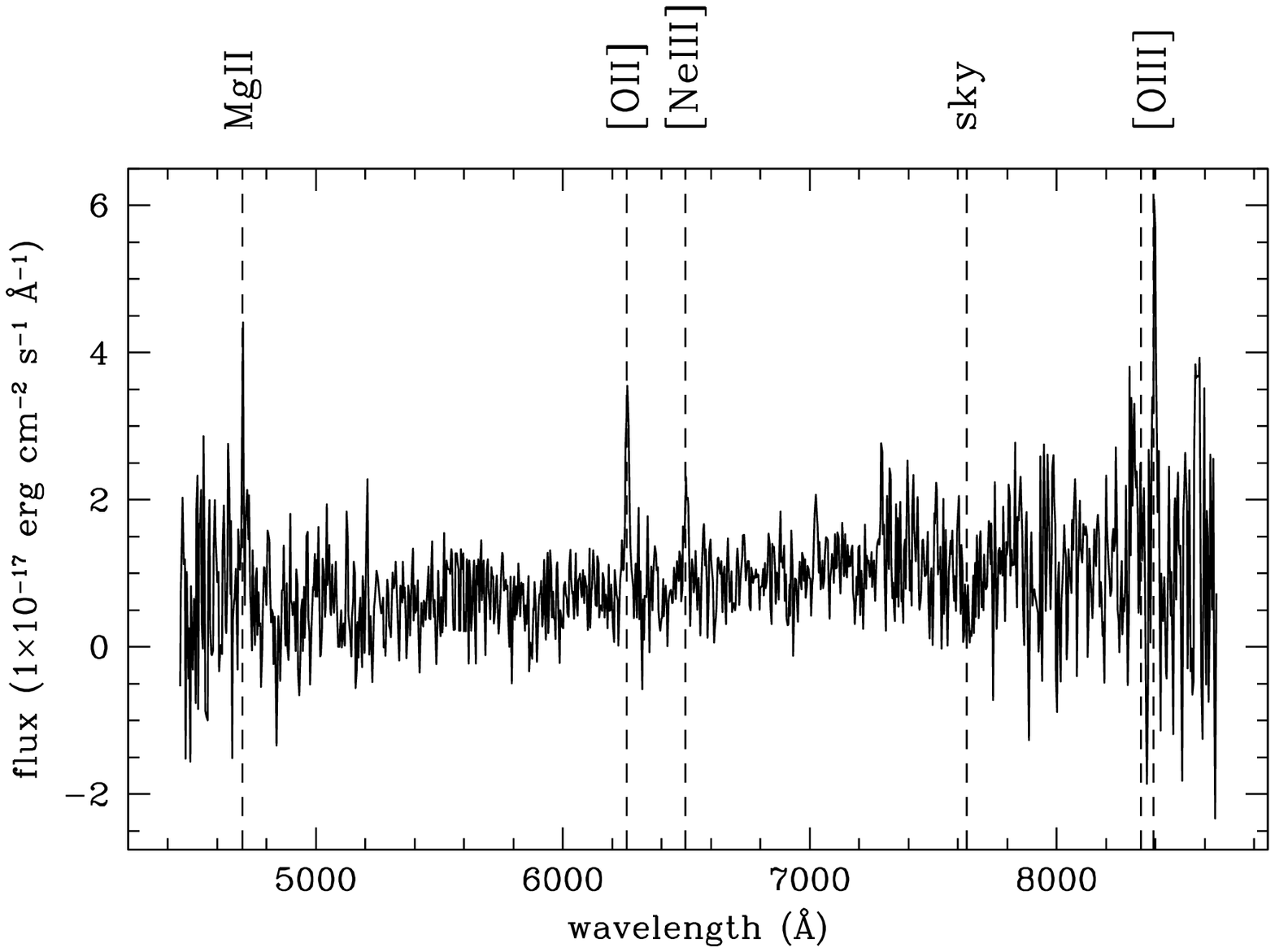}
  \includegraphics[bb=0 0 574 420, width=10cm]{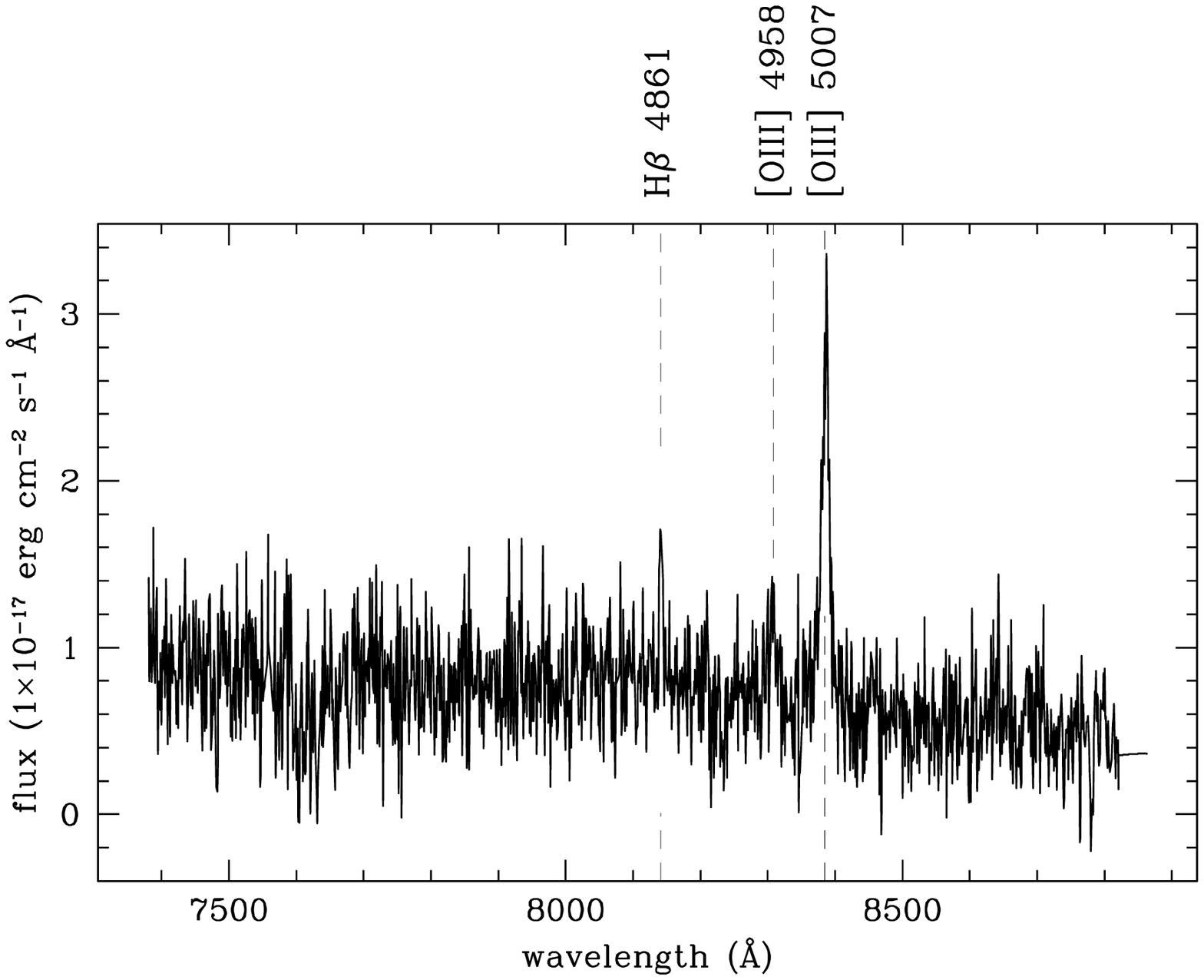}
  \includegraphics[bb=0 0 574 420, width=10cm]{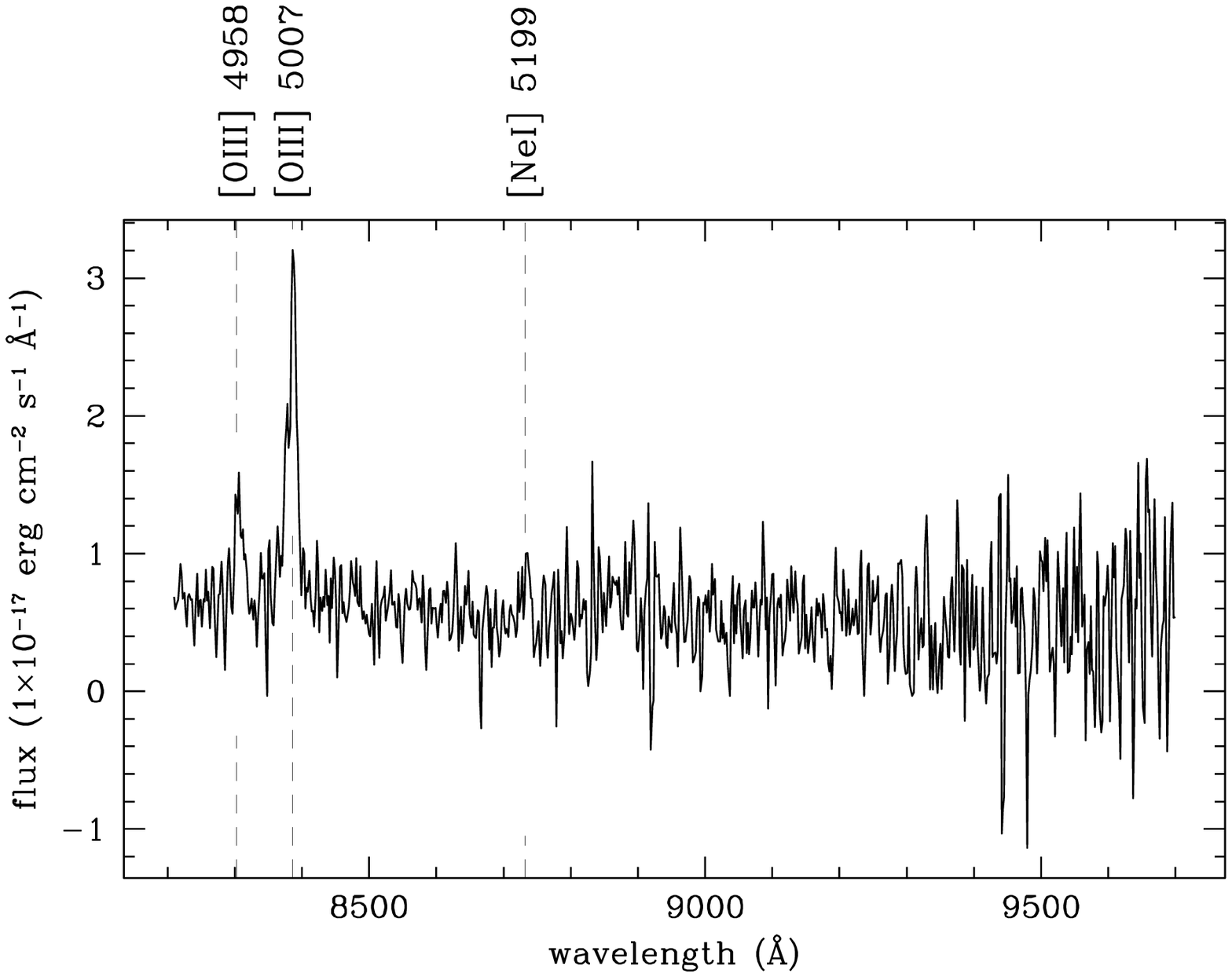}
\vspace{-4cm}
   \caption{NEP\,1640: (top) longslit spectrum of RXJ\,1724.9$+$6636, QSO\,2
    candidate at z$=$0.6758, obtained with the University of Hawai$'$i 2.2m
    telescope; (center) longslit spectrum of the same object taken at the 
    TNG 3.6m telescope with the High Resolution grism and (bottom) the
    Medium Resolution grism. The dashed lines indicate the positions of the 
    emission lines at the object redshift. Wavelengths of atmospheric    
    absorption (or bad sky subtracion) are also indicated.}
\label{fig2}
\end{figure}

The optical spectra of the NEP QSO\,2 candidates, taken at the UH 2.2m 
telescope of the Mauna Kea Observatory with the Wide Field Grism Spectrograph,
are shown  in Fig.~\ref{fig1} and Fig.~\ref{fig2} for two of the three QSO\,2 
candidates. These spectra are the original discovery  spectra and have
relatively low signal to noise ratio. 
However, the presence of  [NeIII]  is indicative of high-ionization line  
objects.  New spectroscopic observations were carried out at the TNG 3.6m 
telescope located at Roque de Los Muchachos Observatory in the Canary islands
on 2004, May  19, and June 19, 22 and 23.
The new spectra were acquired to obtain higher resolution and larger 
wavelength coverage in the optical-IR regime in order to confirm that 
the permitted lines are narrow, including H$\alpha$ when present in 
the observed wavelength range.

\medskip\noindent
We used the low resolution spectrograph DOLORES (Device Optimized 
for the LOw RESolution). The Dolores camera was operated at the  f/3.2 focus. 
The camera is equipped with a Loral thinned and back-illuminated CCD 
with 2048 x 2048, 15$\mu$ pixels which provided a derived scale of 
0.275 arcsec pixels$^{-1}$, and a field of view of about 9.4 x 9.4 arcmin.
A combination of medium and high resolution spectra were obtained centered in
the blue or red/infrared part of the spectrum, according to the redshift 
of each QSO\,2 candidate. We used the MR-I grism for NEP\,1239, the 
highest redshift object at z$=$0.85 for a total exposure of 2.5 hours 
divided into three exposures of 3000s each. 
We requested also an infrared spectrum to check for the presence of 
H$\alpha$.  This spectrum however was never acquired due to the faintness 
of the object which was not visible when the observation was performed.
We may speculate that the object probably varied (as  is common for AGN)
at the time of the TNG observations. In fact NEP\,1239 is the only object
with a  {\em ROSAT} PSPC flux  (see Tables \ref{tab1} and \ref{tab4}) 
which is about twice the value of the  XMM-{\em Newton} flux 
measured almost simultaneously to 
the TNG observations. For NEP\,1640 (z$=$0.6758) the HR-I and 
MR-I grisms were used for a total exposure of about 2h divided into
4500s for the HR-I and 3000s for the MR-I, while for  NEP\,2131 (the 
lowest redshift object at z$=$0.358) the MR-B (700s) and MR-I 
(1100s) grisms were used. The HR-I grism provides a 
pixel scale in spectroscopic mode of 0.8 \AA\ pixels$^{-1}$.
The slitlet width of 1$''$  provided a spectral resolution of 
$\sim$3 \AA\ FWHM. The wavelength coverage was  approximately 
7360--8900 \AA\ . The  MR-I grism gives  a pixel scale of 1.8 \AA\ 
pixels$^{-1}$, a spectral resolution of $\sim$7 \AA\ FWHM, and a
wavelength coverage of approximately 8400--12,000 \AA\, which in reality 
turned out to be only up to 10,000 \AA\ . The  MR-B grism provides  
a pixel scale of 1.7 \AA\ pixels$^{-1}$, a spectral resolution 
of $\sim$6 \AA\ FWHM, and a wavelength coverage of approximately 
3500--7000 \AA\ . All the values are for 1\arcsec slit used.

\begin{figure}
   \centering
   \includegraphics[bb=0 100 574 654, width=14cm]{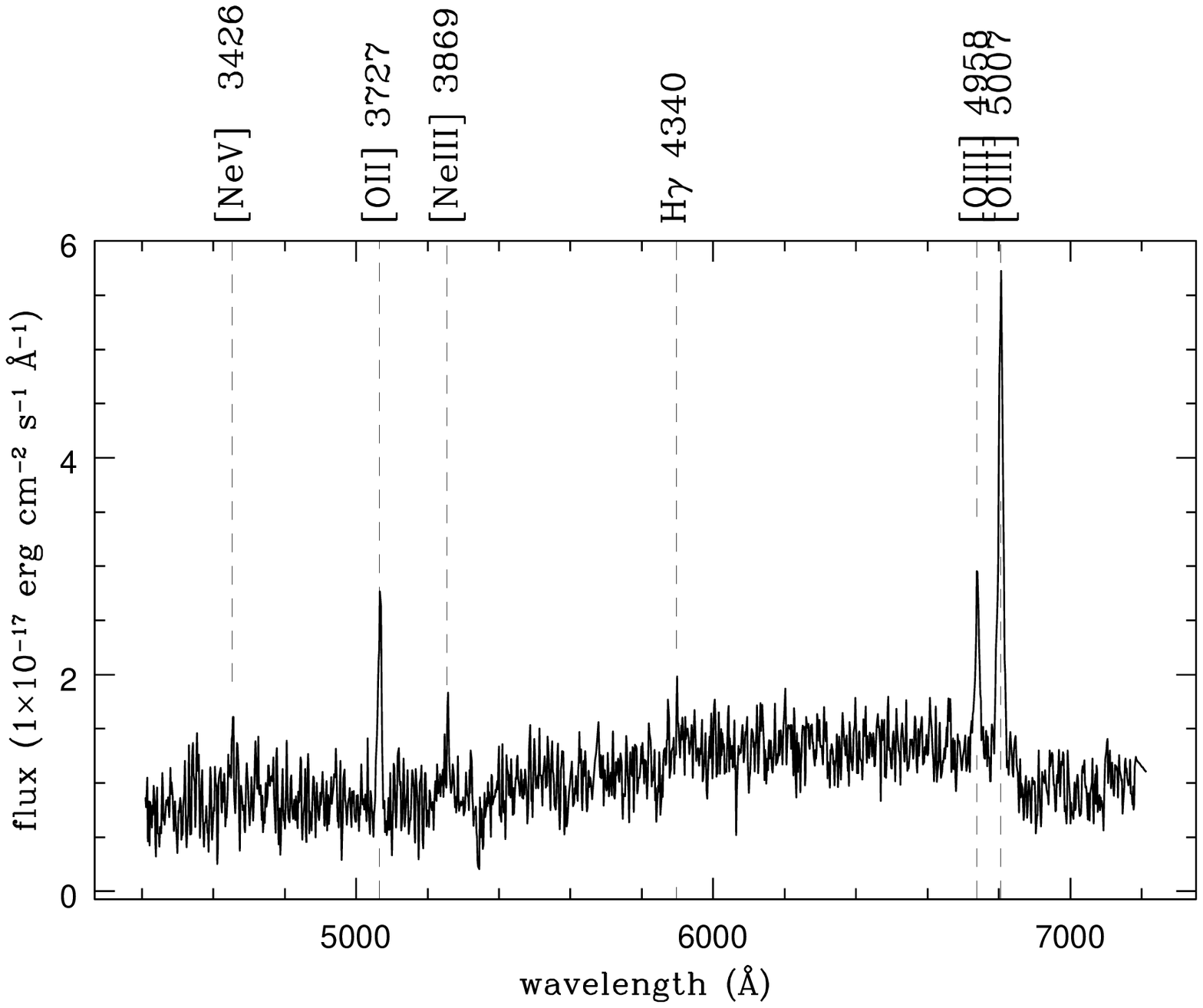}
   \includegraphics[bb=0 100 574 584, width=14cm]{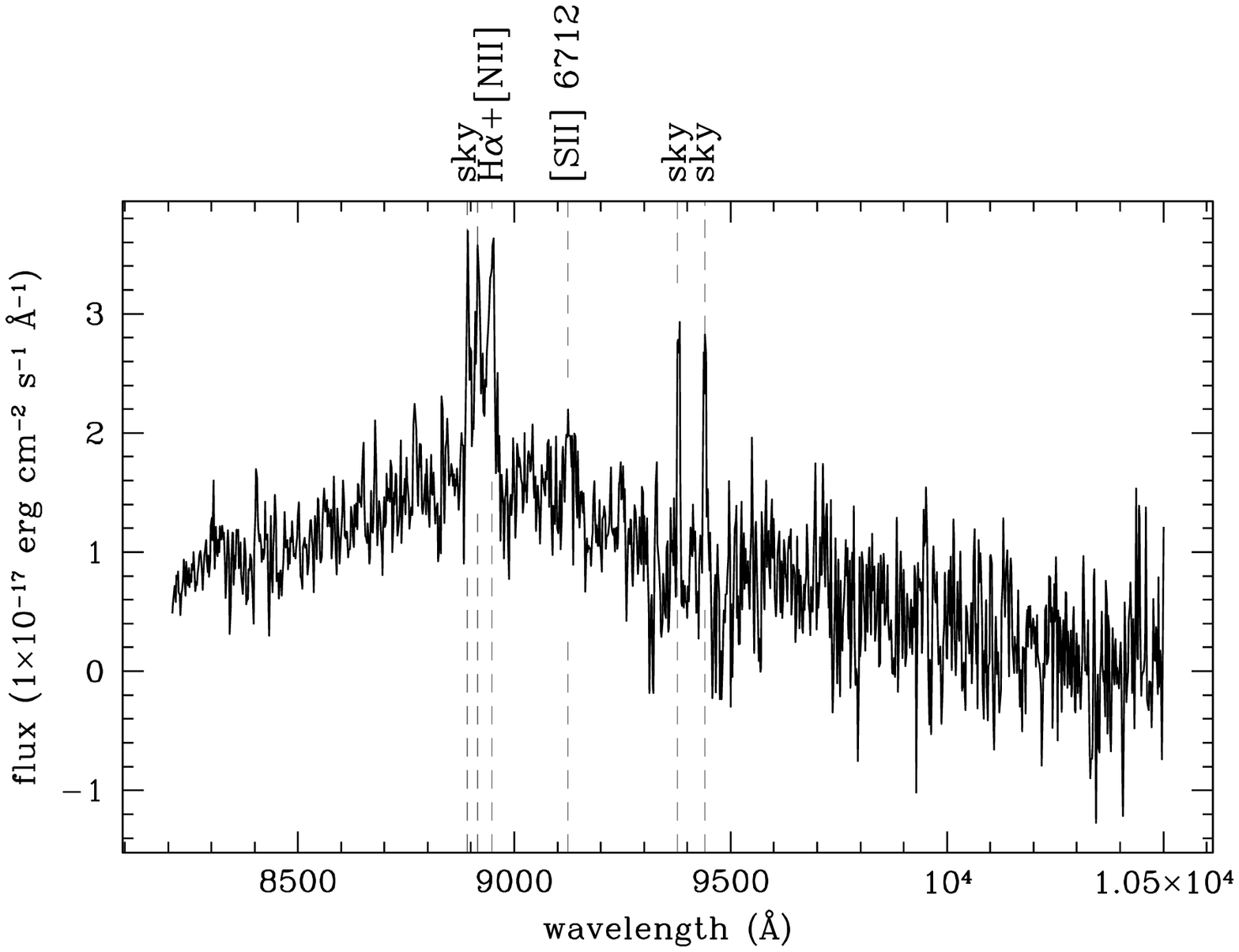}
   \vspace{-3cm}
   \caption{NEP\,2131: longslit spectrum of RX\,J1737.0$+$6601, QSO\,2
     candidate at z$=$0.358, obtained with the TNG 3.6m telescope and the 
     Medium Resolution grism in the bluer (top) and redder (bottom) 
     wavelength region. The dashed lines indicate the positions of 
     the emission lines at the object redshift. Wavelengths 
     of bad sky subtraction are also indicated.}
\label{fig3}
\end{figure}

\medskip\noindent
The optical data were analyzed using  standard IRAF\footnote{{\em
http://iraf.noao.edu\/}} reduction packages. The  redshifts measured 
at the TNG 3.6m for the three objects are in perfect agreement with the 
previously determined redshifts from the discovery spectra as can be 
seen in Table~\ref{tab1}.
The TNG spectra for the three objects are shown in Fig.~\ref{fig1} (bottom),
Fig.~\ref{fig2} (center and bottom) and Fig.~\ref{fig3}. The following 
properties, listed also in  Table~\ref{tab2}, can be derived from the 
optical spectra of the three objects:

\begin{itemize}
\item
NEP\,1239: the TNG redder wavelength spectrum (see Fig.~\ref{fig1}) for this
object at z$=$0.85 nicely complements the UH spectrum and shows narrow 
H$\beta$ (FWHM$=$453 km s$^{-1}$) and the [OIII] doublet emission lines 
(FWHM$=$458 km s$^{-1}$ for  [OIII]\,$\lambda$4958 \AA\ ; 
FWHM$=$457 km s$^{-1}$ for 
[OIII]\,$\lambda$5007 \AA\ ). The H$\alpha$  region is outside the wavelength 
range of the TNG spectrum, so unfortunately no information is available. 
The log ([OIII]\,$\lambda$5007 / H$\beta$) is $=$0.79, which places it in
the Seyfert\,2 region of the spectral classification plots of \citet{vo87}.

\vskip 0.5truecm
\item
NEP\,1640: in the UH 2.2m spectrum (Fig.~\ref{fig2}-top) four narrow 
emission lines were detected: MgII, [OII], [NeIII] and [OIII] respectively. 
The two TNG spectra are shown in the same figure (the High Resolution 
spectrum  in the middle, and the Medium Resolution spectrum at the bottom). 
The H$\beta$ emission line is visible in the HR spectrum with a measured  
FWHM$=$270 km  s$^{-1}$, while [NeI] is better visible in the bottom 
spectrum of  Fig.~\ref{fig2}. The respective rest-frame 
full widths half maxima are FWHM$=$480 km 
s$^{-1}$ for [OIII]\,$\lambda$4958 \AA\ , FWHM$=$623 km s$^{-1}$ for 
[OIII]\,$\lambda$5007 \AA\  and FWHM$=$608 km s$^{-1}$ for 
[NeI]\,$\lambda$5199 \AA\ . H$\alpha$ for this object at z$=$0.6758 
would fall at $\sim$11,000 \AA\ , outside the range covered by this 
spectrum.  The log ([OIII]\,$\lambda$5007 / H$\beta$) is $=$0.94, even
higher than the NEP\,1239, which can be again classified from the optical
spectrum as a Seyfert\,2 object.

\vskip 0.5truecm
\item
NEP\,2131: only for this object at z$=$0.358 we were able to detect 
H$\alpha$. No UH spectrum was taken. The Multiple Mirror telescope spectrum
obtained for us in July 97 by John Huchra is not given here since it is very 
similar to the TNG MR-B spectrum shown in Fig.~\ref{fig3} (top). The lines 
have the 
following widths: [NeV] has FWHM$=$819 km s$^{-1}$, [OII] has FWHM$=$801 km 
s$^{-1}$, [NeIII] has FWHM$=$460 km s$^{-1}$, H$\gamma$ has FWHM$=$633 km 
s$^{-1}$,  [OIII]\,$\lambda$4958 \AA\ has FWHM$=$741 km s$^{-1}$ and  
[OIII]\,$\lambda$5007 \AA\ has  FWHM$=$707 km s$^{-1}$. 
H$\beta$ is not detected, an upper limit gives a flux less 
than $\sim$2.2$\times10^{-16}$ erg cm$^{-2}$ s$^{-1}$ \AA\ $^{-1}$. 
The log ([OIII]\,$\lambda$5007 / H$\beta$) is $>$1.22. 
H$\alpha$ is visible in the TNG MR-I spectrum (bottom in Fig.~\ref{fig3}). 
There are several sky lines which did not subtract well and which are 
indicated as ``sky''. The FWHM of H$\alpha$ is 726 km s$^{-1}$ while the FWHM 
of [NII] is 804 km s$^{-1}$. The log ([NII]\,$\lambda$6583/H$\alpha$) is 
equal to 0.09, while log ([SII]\,($\lambda$6712+6731)/H$\alpha$) is equal 
to -0.5. The line ratios place this object in the Seyfert\,2 region of the 
diagnostic  plots by \cite{vo87} or according to the recent revision of 
\cite{kauf03}.

\end{itemize} 
\smallskip\noindent
Thus all the emission lines detected in the three QSO\,2 candidates are 
narrow lines, all with a FWHM less than 1000 km s$^{-1}$, well within the 
cut-off  broadness of 2000 km s$^{-1}$ set for classification purposes in the 
NEP survey.

\tabcolsep 0.3cm
\begin{center}
\begin{table*}
\caption{Journal of XMM-{\em Newton} Observations}
\label{tab3}
\begin{center}
\begin{tabular}[h]{l r l r r r r r }
\hline
\hline
NEP id \# & Date & Instrument & Total  &  Flare-free & Total & Flare-free \\
         &   &  & Net Counts  & Net Counts & Exp. Time   & Exp. Time  \\
         &   &  & (0.5-10 keV) & (0.5-10 keV) & sec & sec \\
\hline
\\		
NEP\,1239 & 06/27/2004 &  MOS1 & 414.1$\pm$29.6 & 105.8$\pm$12.3 & 32,602 &   11,755   \\ 
      &  & MOS2 &  485.8$\pm$33.6 &  150.4$\pm$14.1  & 32,575  &   12,165  \\
  &     &  pn &  1238.0$\pm$77.0 & 299.8$\pm$18.8  & 24,761   &   5,837   \\
NEP\,1640 & 03/11/2004 & MOS1 & 416.4$\pm$22.2 &  217.3$\pm$15.7 & 29,372 & 13,233   \\
     &  and 04/14/2004 & MOS2 & 427.2$\pm$23.4 &  243.6$\pm$16.1 & 29,385 & 15,093   \\
      &               &   pn & 1339.9$\pm$45.4 &  412.4$\pm$20.7 & 26,612  & 6,022   \\
NEP\,2131 &  05/14/2004 & MOS1 & 1063.9$\pm$41.5 &  685.0$\pm$27.3 &  30,955 & 21,433  \\
          &    & MOS2 &  960.4$\pm$39.7 &  661.7$\pm$26.7 & 31,061 & 21,533 &  \\
         &     & pn  &  2387.3$\pm$79.5 & 1892.9$\pm$45.6 &  24,063  & 16,905 \\
\\
\hline  
\end{tabular}
\end{center}
\end{table*}
\end{center}

\subsection{XMM-{\em Newton} Data}

XMM-{\it Newton} \citep{jan01} observed NEP\,1239, NEP\,1640 and NEP\,2131
as part of the GO program in four epochs, March through June 2004, 
with the  European Photon Imaging Camera (EPIC) pn 
\citep{stru01} and with the EPIC  MOS CCD arrays \citep{tur01}.
The journal of the observations is reported in Table~\ref{tab3}.
The pn and MOS instruments were operating in full-frame mode with the
thin filter (thick for NEP 1640) applied.  Event files produced from 
the standard pipeline processing have been examined to recognize
high background time intervals (using the version 6.1.0 of the Science 
Analysis Software, SAS, and the latest calibration files released by the 
EPIC team). Only events corresponding to pattern 0-12 for MOS and pattern 
0-4 for pn have been  used\footnote{see the XMM-{\it Newton} Users' 
Handbook \\
{\it http://xmm.vilspa.esa.es/external/xmm\_user\_support/documentation/uhb/XMM\_UHB.htm}}.
Good time intervals were selected by applying thresholds of 0.35 counts 
s$^{-1}$ in the MOS and 1 counts s$^{-1}$ in the pn to the photons at 
energies greater than 10 keV. At these higher energies counts come mostly 
from background. 
The net exposure times, before data cleaning, are of the order of 24-32ks 
as listed in Table~\ref{tab3}. These exposure times reduce to 5-21ks after
data cleaning thus reducing dramatically the count statistics. 
The available statistics after cleaning becomes so poor that we decided 
to retain time intervals of background flares. Given the point-like 
morphology of the sources the total background count rate under their
extraction regions is still not very high even in the presence of 
possible flares. On the other hand the improvement in source statistics 
is much higher than the drawback of having a high noise. 
In any case, we have performed analysis on both the total and flare-free
datasets and we find results consistent within the errorbars for
all parameters. Net counts and exposure times in seconds, before and 
after cleaning (Total and Flare-free), are reported in Table~\ref{tab3}. 
X-ray source counts were extracted in circular regions centered on the
source with radius 30$^{\prime\prime}$ (except for 
NEP\,1640 where a smaller 20$^{\prime\prime}$ radius was used due to 
the presence of other sources close to the target) and  background 
counts were extracted in a nearby source-free circular region 
using a radius of 100$^{\prime\prime}$. For only one source, NEP\,1640, 
the X-ray observations were performed in two diferent epochs.
The datasets were independently analyzed. Both source and background 
spectra have been summed, while response matrices and auxiliary
files are the average of the two exposures, weighted by the different
exposure times. Response matrices at the position of the target source  
have  been generated using the SAS tasks {\it arfgen} and {\it rmfgen}.
The two MOS spectra are summed to improve the statistics.
Since the resulting relative normalization between the MOS and pn instruments
is always within 10\% we give only fluxes for the MOS.

\subsection{Spectral Analysis}

We used XSPEC (version 11.3.1) for spectral modelling. Spectral bins below 
0.5 keV were excluded because of uncertain calibration, and those above 
6 (10) keV where excluded for MOS (pn) because of the low statistical 
significance. Spectra were binned 
to have at least 20 counts and at least a significance of 4$\sigma$ in each 
bin.  MOS and pn spectra were jointly fitted  with independent 
relative normalizations to account for possible intercalibration differences 
or mismatches. The spectra are well fitted by a single power law model with 
low energy absorption fixed at the Galactic value along the line of sight
to each object \citep{elv94,star92} (see Table~\ref{tab1}). 
Results are summarized in Table~\ref{tab4}. The second column gives the 
values of the power low slope $\Gamma$ when the N$_{\rm H}$ is fixed to the
Galactic values of Table~\ref{tab1}, while the third column  indicates 
the range of additional intrinsic N$_{\rm H}$ when it is left free to vary. The following columns 
list the $\chi^2_{\nu}$ values and degrees of freedom, null hypothesis 
probabilities, XMM-{\em Newton} fluxes and luminosities in different 
energy bands. The X-ray spectra  are shown in figures 4 through 6. 
Taken at face value the X-ray spectra 
are not consistent with an obscuration scenario, that is the expected 
significant absorption at low energies, typical of QSO\,2 objects, 
is not seen in any of the three XMM spectra. This is very surprising since 
the three  objects all have the characteristics of QSO\,2, that is narrow 
optical  emission lines and X-ray luminosities at the 10$^{44}$ erg s$^{-1}$ 
level or higher.
\begin{center}
\begin{table*}
\caption{Spectral Analysis Results}
\label{tab4}
\begin{center}
\begin{tabular}[h]{l c c c c  c c c}
\hline
\hline
Name   & $\Gamma$ & N$_{\rm H}$ & $\chi^2_{\nu}$/dof & P & f$_{0.5-2 keV}$ & f$_{2-10 keV}$ & L$_{2-10 keV}$ \\
     &   & 10$^{20}$ cm$^{-2}$ & & \%  & 10$^{-13}$ erg cm$^{-2}$ s$^{-1}$ & 10$^{-13}$ erg cm$^{-2}$ s$^{-1}$ & 10$^{44}$ erg s$^{-1}$ \\
\hline
\\
NEP\,1239 & 1.76$^{+0.10}_{-0.10}$ & $< 9$ & 1.03/51 & 40 & 0.61$\pm 0.1$ & 1.01$\pm 0.1$ & 3.1$\pm 0.3$ \\
         &    & & & & \\
NEP\,1640 & 1.88$^{+0.09}_{-0.09}$ & $3-12$ & 0.88/72 & 75 & 0.82$\pm 0.1$ & 1.14$\pm 0.2$ & 2.2$\pm 0.2$ \\
         &    & & & & \\
NEP\,2131 & 1.50$^{+0.06}_{-0.06}$ & $<2 $ & 1.02/143 & 41 & 1.39$\pm 0.1$ & 3.4$\pm 0.3$ & 1.3$\pm 0.1$ \\
         &    & & & & \\
\hline 
\end{tabular}
\end{center}
\medskip\indent
Unabsorbed fluxes and rest-frame K-corrected luminosities; N$_{\rm H}$ is
the range of total absorption at the quasar redshift, 
when left free to vary. P is the null hypothesis probability.
\end{table*}
\end{center}

\begin{figure}
   \centering
   \includegraphics[width=10cm, angle= -90]{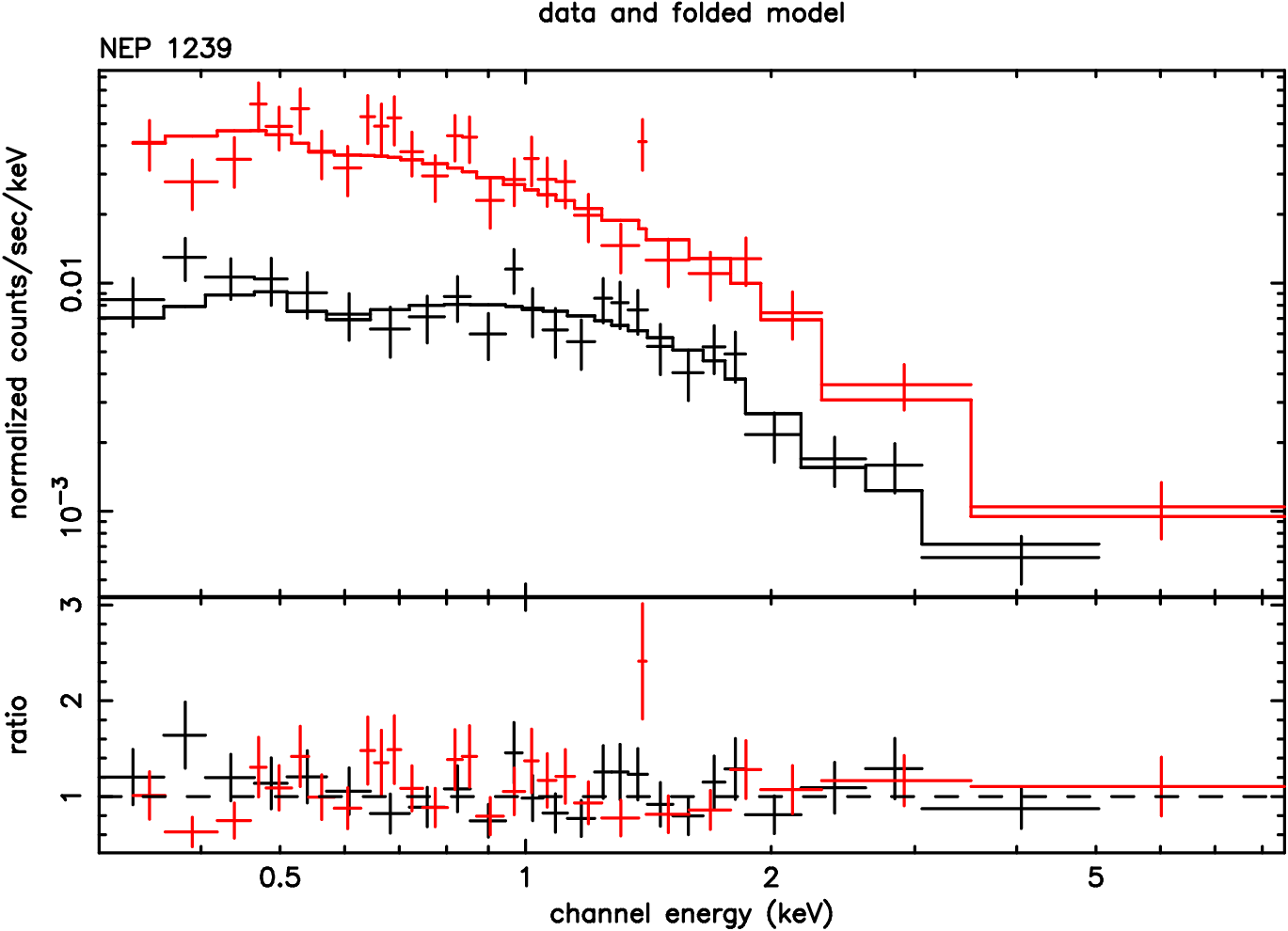}
   \caption{X-ray spectra and model (upper panel) and ratio between the
two (lower panel)
for NEP\,1239.}
\label{fig4}
\end{figure}

\medskip\noindent
The X-ray spectrum for the first object, NEP\,1239, is given in 
Fig.~\ref{fig4}. The spectrum is well fitted by a power law model with 
$\Gamma=$1.76  and shows no sign of absorption at low energies. 
When left free the total N$_{\rm H}$ is $< 9 \times 10^{20}$ cm$^{-2}$ at the quasar redshift.
Its 0.5--2 keV flux is about half the {\em ROSAT} NEP flux. 
Such variations are not unusual for AGN in the X-ray band \citep{mush93}.
The source probably was  in a bright state at the time of the {\em ROSAT} 
observation thus favouring its detection.  Even at the lower flux 
observed by XMM-{\em  Newton} NEP\,1239 would have made the cut-off
threshold  for inclusion in the sample of QSO\,2 candidates.  The faintness 
of the source was probably the reason why the NICS spectrum could not be 
taken (see Sect. 3.1). The fainter X-ray continuum should give more contrast 
to the presence  of a Fe-line, which is however not detected (see below).

\begin{figure}
   \centering
   \includegraphics[width=10cm, angle= -90]{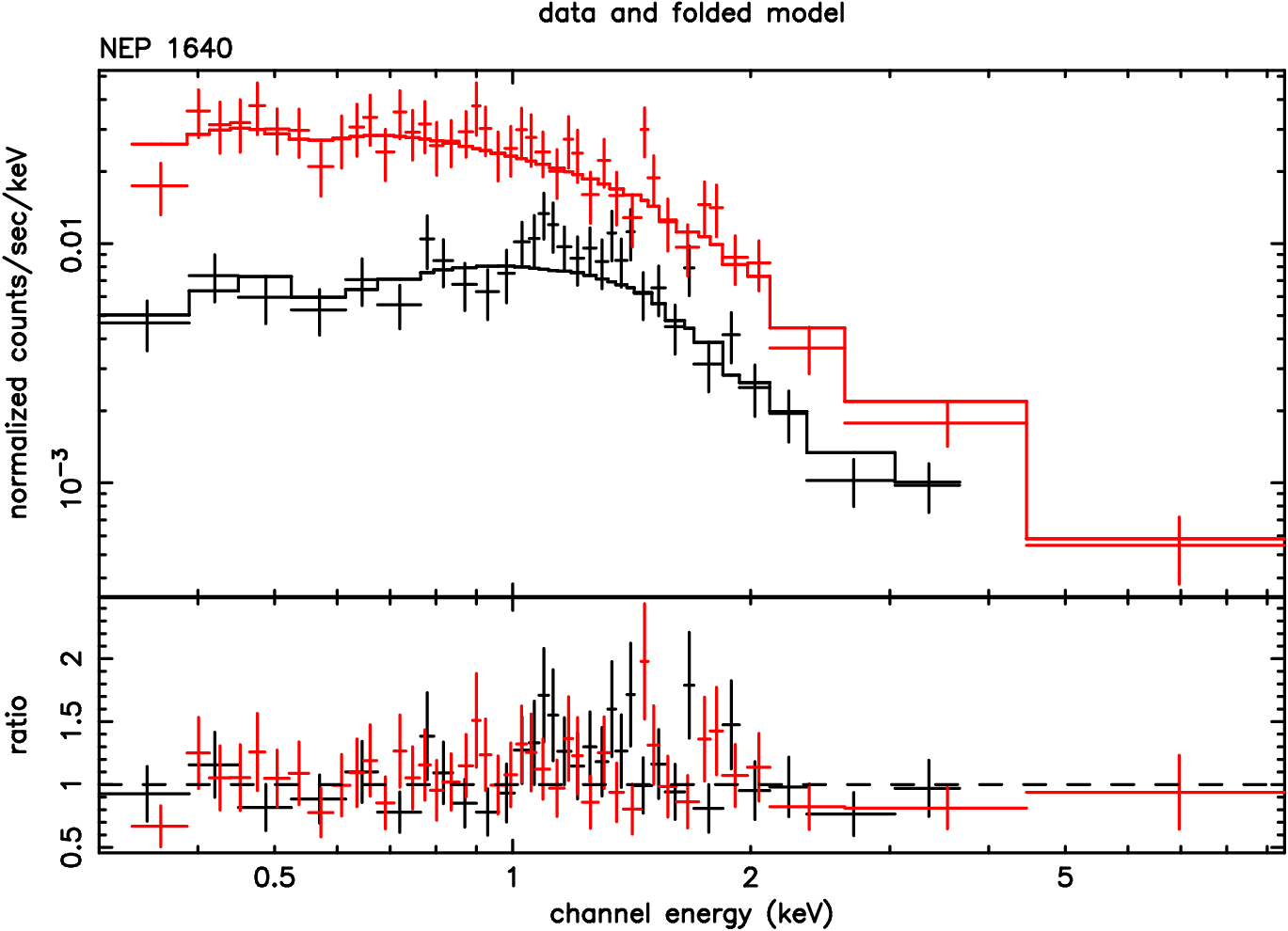}
   \caption{X-ray spectra and model (upper panel) and ratio between the
two (lower panel)
for NEP\,1640.}
\label{fig5}
\end{figure}

\medskip\noindent
The spectrum of NEP\,1640 (Fig.~\ref{fig5}) is again well fit by a
power law model with $\Gamma=$1.88, with Galactic absorption.
When left free to vary the total N$_{\rm H}$ ranges between $ 3 - 12 \times 
10^{20}$ cm$^{-2}$ at the quasar redshift. 

\begin{figure}
   \centering
   \includegraphics[width=10cm, angle= -90]{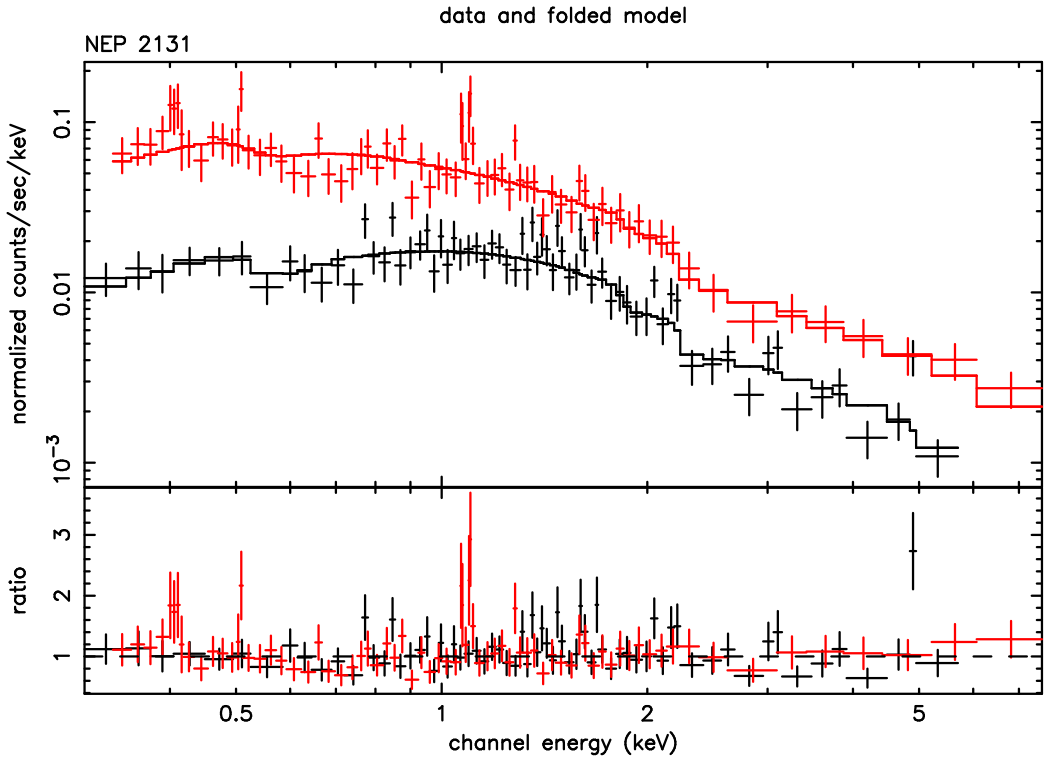}
   \caption{X-ray spectra and model (upper panel) and ratio between the
two (lower panel)
for NEP\,2131.}
\label{fig6}
\end{figure}

\begin{figure}
   \centering
   \includegraphics[width=10cm, angle= -90]{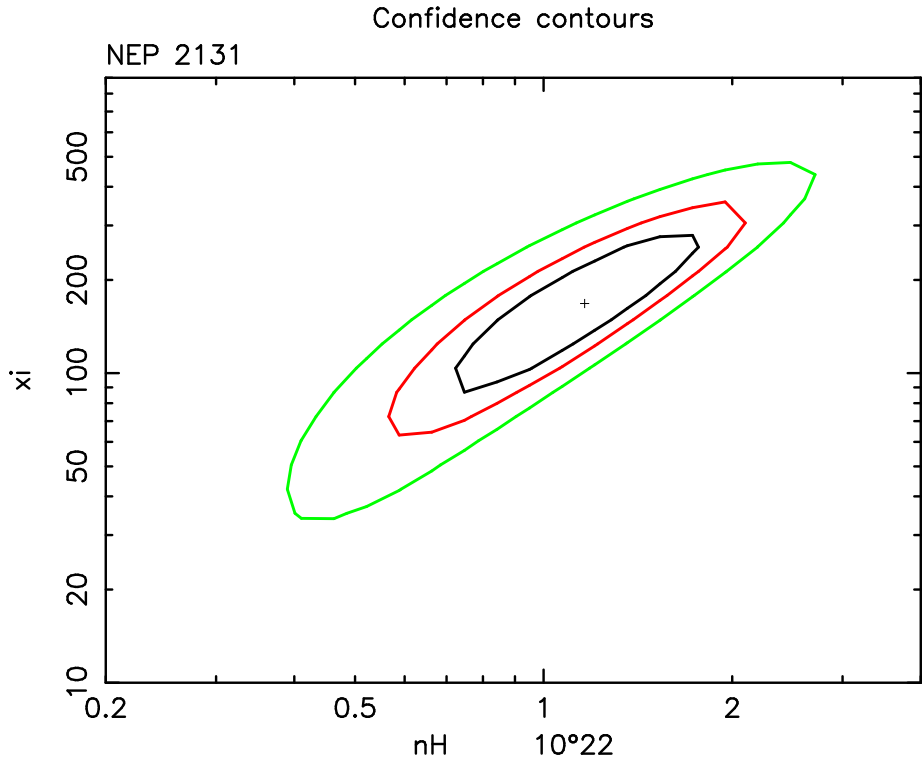}
   \caption{Contour plot of the two free parameters in {\sl absori + power 
law}, N$_{\rm H}$ and $\xi$, for NEP\,2131. The power law slope is fixed 
to the best fit value of $\Gamma = 1.7$.}
\label{fig7}
\end{figure}

\medskip\noindent
NEP\,2131 has a rather flat spectrum (see Fig.~\ref{fig6}). The fit is good, 
from a statistical point of view (see Table \ref{tab4}). When left as a 
free parameter the total N$_{\rm H}$ is $< 2 \times 10^{20}$ 
cm$^{-2}$ at the quasar redshift. 
We can nonetheless investigate the possibility that the flatness is 
only apparent and indicative of absorption. 
A good fit is obtained by fixing the power law slope to the canonical 
$\Gamma = 1.7$ value, with an absorption in the range (1.1 -- 2.5) $\times$ 
10$^{21}$ cm$^{-2}$ (in the source rest frame). The residual low energy excess 
is well described by a thermal plasma (model {\sl mekal} in XSPEC) with 
kT $=$ 0.16$^{+0.03}_{-0.03}$ keV.
The standard explanation for the soft X-ray excess is thermal emission 
originating directly from the hot inner accretion disk \citep{ms82}.   
However for a sample of low redshift quasars in which both M$_{BH}$ and 
$\dot{m}$  are measured \citep{por04}, the expected maximum
  temperature in the geometrically thin, optically thick accretion disk
  scenario is of the order of tens of eV,
while the measured 
temperatures in the soft excess are of a few hundred eV. Compton 
upscattering of soft (EUV) disk photons in a hot plasma (the corona
above the disk) is therefore a more realistic scenario for these objects.
The measured temperature for the possibile soft excess in NEP\,2131
with the same {\sl diskpn} model as in \cite{por04} yields 
kT $=$ (7.5$\pm 1.5) \times 10^{-2}$ keV, lower than what is measured for 
the low redshift quasars, but not enough to be explained by the accretion 
disk directly. 
An alternative scenario implies the presence of a ionized, high column
density absorber. We have tested it by adding an ionized warm absorber
({\sl absori} in XSPEC) to the power law fit. The formal fit gives
an N$_{\rm H} \sim 1. \times 10^{22}$  and log$\xi = 2.2$. The contour
plot of the error uncertainty of the 2 parameters is shown in Fig.~\ref{fig7}.
The formal best fit values are similar to those of the Sy 1 NGC 3783 
\citep{net03}, which has a much higher statistics.
However, the addition of the new component affects the spectrum only below
$\sim 1$ keV. In this energy range the addition of a new component is 
significant only at 97\% confidence (according to the F-test). 

\medskip
\medskip\noindent
For the three sources, if the observed spectrum is due to reflection over a 
neutral medium, we would expect a prominent iron K$\alpha$ line at around 
6.4 keV rest-frame \citep[see e.g.][]{mbf96}. 
No line is evident in any of the three objects
spectrum. However, given the relatively poor statistics, we fitted a 
narrow 6.4 keV line to each spectrum and derived the following equivalent
width (EW) upper  limits. For NEP\,1239 we derive EW$ <$ 250 eV, 
for NEP\,1640 EW $<$ 190 eV, for NEP\,2131 EW $<$240 eV. 
To better investigate the 
possibile presence of a Fe line we also used the software FLEX 
\citep[described in][]{mac04,bra05} which searches for high S/N excesses
in unbinned spectra.  Even with this ad-hoc procedure no 
excess due to a possible Fe line at the source redshift was found in 
any of the three objects.

\section{Discussion and Conclusions}

From the optical point of view the three NEP objects all have
the properties of Seyfert\,2 objects both in the UH 2.2m and in
the higher resolution TNG 3.6m spectra. 
The absence of a broad component in the H$\beta$ or H$\alpha$ emission
lines implies an extinction of at least a magnitude which for standard 
galactic relationship, 
N$_{\rm H}$/A$_V = 1.79 \times 10^{21}$ cm$^{-2}$ \citep{ps95},
corresponds to N$_{\rm H} \magcir 1 \times 10^{21}$ cm$^{-2}$.
We cannot set a more stringent limit from optical data
since the broad components of H$\alpha$ and H$\beta$ are not
detected. However this lower limit is larger than the upper limit derived
from simple modeling of the X-ray data. Within the Unified Model 
\citep{ant93} we assume that the BLRs in these objects are obscured by 
some kind of medium, usually referred to as the ``torus''.
The optical light is mainly absorbed by dust, and the X-ray radiation
by gas. The torus contains both media and therefore it absorbs both the 
optical and the X-ray emission. This relation between optical and X-ray 
absorption is supported by observational evidence \citep[see e.g.][]{cac04},   
where optical and X-ray  classification for obscuration agree, that is the 
optical and X-ray  measured absorptions seem to be linked, or at least 
present in the same objects. 
There are a few known counter-examples \citep[e.g.][]{bru03,pag03,np94}  
in which the X-ray derived absorption is higher than  what is seen
in the optical. This situation can have different explanations.
It could be due to a different dust-to-gas ratio, to the presence 
of ``large'' dust grains \citep{mai01}, or to the dust being swept
out by the ionizing radiation. 

\medskip\noindent
In the three objects under study here 
we have a situation where an absorber masks the radiation coming from 
an external zone, the BLR, but not the one from the inner zone, the nucleus.
The X-ray spectrum is in fact well fitted by a canonical power law
slope ($\Gamma \sim 1.7$) with no requirement for additional intrinsic
absorption at the source. Only if one assumes a more complex model,
which is not statistically required by the data (as in NEP\,2131 where 
the model has two additions: the absorption at low energies and an extra 
thermal component in the same energy range to account for the residuals), 
then a larger value for the intrinsic absorption is required.
These three objects also differ from other objects, like NGC\,4698, 
which is a peculiar Seyfert\,2  with no intrinsic absorption,
possibly the end product of a merger event \citep[see e.g.][]{ber99},
because they have high X-ray luminosities. \cite{pap01}  describe 
NGC\,4698 as either a dusty warm absorber, or as an object with intrinsic 
lack of Broad Line clouds due to its  low X-ray luminosity (i.e. lack of 
ionizing radiation to illuminate the clouds). 
\cite{gz03} recompute the source luminosity to avoid contamination
of other sources and reach the same conclusion of a lack of Broad
Line Regions possibly due to low accretion rate. The three NEP objects 
all have luminosities at L$_X \sim 10^{44}$ erg cm$^{-2}$ s$^{-1}$  level 
thus making this  explanation even less viable.

\medskip\noindent
By using a sample of optically selected Seyfert\,2,  \cite{pb02}
estimate that the fraction of such galaxies which is Compton thin (N$_H \leq
10^{22}$ cm$^{-2}$) is of the order of 10--20\%. They propose the following
explanations: either the BLRs are covered by a dusty obscuring material 
(dust patches, dust lanes, HII regions) or the BLRs are absent, weak or 
faded away. This interpretation is supported by the low (L$_X \sim 10^{41-42}$
erg s$^{-1}$) luminosity of the nuclei. In addition the percentage of Compton 
thin objects seems to decrease with increasing luminosity, so this 
interpretation it's not  useful for our case. At higher luminosity,
L$_X = 2 \times 10^{43}$ erg s$^{-1}$, \cite{cac04} report a Sy\,1.9 with 
no  absorption  representing, albeit large uncertainties, a fraction in 
their sample of   12\%, similar to that found by  \citet{pb02}.
The presence of BLR clouds has been linked to the accretion rate 
\citep{nic00} in such a way that there is an upper limit
to the width of the lines corresponding to a minimum accretion rate dictated 
by the conditions of a \citet{ss73} accretion disk.
Thus low luminosity, low accretion rate nuclei cannot have Broad Emission
Lines. The case for the present QSO2 is however more complex: they might
have a very low accretion rate or, conversely, a very high one that
would give origin to a BLR very far away from the nucleus and 
therefore in a region of low velocities that do not widen the lines.

\medskip\noindent
There is always the possibility that the torus in the three NEP  objects 
is Compton thick. The X-ray component would then be the reflection off the 
far side of the torus, that mantains the same slope as the direct component 
which is however completely blocked in the observed energy range. 
However, the F$_X$/F$_{[OIII]}$ ratio 
for the three NEP objects has values
consistent with ``normal'' type 1 objects,  albeit within the calibration 
uncertainties of the optical spectra (for the use of the   
F$_X$/F$_{[OIII]}$ as a diagnostic to find Compton thick objects, 
\citep[see ][]{mai98, pb02}.
Furthermore, we would expect in the reflection spectra a prominent line
(EW $\sim$ 1 keV) from neutral Fe \citep{mat01}. This line is not observed 
in any of the three objects. A further obstacle to the Compton
thick interpretation is the high L$_X$ observed. If this is just a fraction 
(0.1-10\%)  of the total power, the intrinsic luminosity would be very 
high (10$^{45-47}$ erg s$^{-1}$). The spatial/volume density 
would need to be compared with the density of known AGN, that
decrease accordingly by 2-5 orders of magnitude for increasing luminosity 
\citep[see e.g. the XLF in][]{miy00}. 
The NEP QSO2 would represent a higher and higher fraction, even exceeding
by orders of magnitude the density of known AGN.

\medskip\noindent
The most similar object we have found in literature is H1320$+$551 
\citep{bar03}, which is a Seyfert\,1.8/1.9. We do not have the same 
spectral resolution and S/N as the above authors, so we cannot 
exclude that also our  candidates have a broad base to the Balmer lines,
but the overall shape is the same. \cite{bar03} suggest that the narrow 
line region is internally reddened in H1320+551, but with a small covering 
factor over the nuclear emission, and that the Balmer decrement of the BLR 
is an instrisic property rather than caused by reddening or absorption. 
H1320$+$551 has a weak Fe  K$\alpha$ detection (EW $=$ 400eV) too faint 
for the source to be  Compton-thick. A similar example is 
XBSJ031146.1-550702 classified as Sy\,1.9 by  \cite{cac04}. Its X-ray 
emission does not show significant absorption and is probably Compton-thin.
If one assumes that what is seen in these sources at 
X-ray wavelengths is actually the direct nuclear component then one could 
still keep the Unified Model paradigm by assuming that the intervening gas
is completely ionized, and therefore transparent to the X-ray radiation,
while the dust in the torus absorbs the optical BLR. This in turn implies 
that the accretion is inefficient: moderate accretion rate with low
radiative efficiency in the form of either an Advection Dominated Accretion
Flow \citep[ADAF; see e.g.][]{ny95} or a Radiatively Inefficient Accretion
Flow \citep[RIAF; see e.g.][]{qua03},  and therefore at sub-Eddington 
regimes. To be sub-Eddingtion 
(i.e. L$_{bol}$/L$_{Edd} < 10^{-3}$) implies L$_{Edd} > 10^{48}$ 
erg s$^{-1}$ for these objects and therefore M$_{BH} \geq 10^{10}$ 
M$_{\odot}$. At the moment we cannot estimate the black hole 
mass, but  this would be a crucial information to acquire to 
push the limits of  the Unified Models.

\medskip\noindent
The virtually complete identification rate of  the {\em ROSAT} NEP
survey \citep[99.6\%,][]{gio03} allows us to derive for the first time 
an estimate of the spatial density of such sources in soft X-ray 
sky. The spatial density of the NEP QSO\,2 candidates is 
$2.8 ^{+2.7}_{-1.5} \times 10^{-8} h^3 Mpc^{-3}$ with a density-weighted
average luminosity of  L$_{0.5-2} = 1.3 \times 10^{44}$ erg s$^{-1}$. 
We can compare it with the most recent estimates in literature  
\citep[see discussion in][]{ued03}. 
The NEP point is an order of magnitude below any of the different
redshift bin curves  in Fig. 15 of Ueda et al. 2003. Therefore this
indicates that this class of soft X-ray selected unobscured QSO2, 
if it is indeed a new class, and if the are really unobscured
provides a very small contribution to the total luminosity density.
Since these are most probably objects with no absorption, they
do not make up the missing population of the XRB. This result is not 
completely unexpected since the sources have been selected to be bright 
in the soft X-ray band of {\em ROSAT}, while obscured objects are better 
detected in  hard X-rays where the reflection component becomes dominant, 
and/or the direct component is not absorbed. 
Similar objects might be present in current X-ray surveys. For instance, 
unobscured QSO\,2 candidates might be lurking in hard X-ray surveys like 
the ASCA Medium Sensitivity Survey \citep{aki03}, where the low hydrogen 
column density is inferred by a hardness ratio rather than from a more 
precise X-ray spectrum. See the large crosses at the bottom of Fig. 10a 
in \cite{aki03} which represent AGN with no broad H$\beta$, low N$_{\rm H}$ 
(log N$_{\rm H}$ $<$ 20) but large  X-ray luminosity  
(log L$_{\rm X}$ $>$ 44). Also the XMM-Newton Bright Serendipitous 
Survey \citep{rdc04} contains possibile unobscured QSO2 candidates. 
A large fraction (more than 60\%) of the Narrow Line AGN in this survey
(see Fig. 4a in Della Ceca et al., 2004,  and relative discussion) has 
an hardness ratio 
in the same range as the Broad Line AGN. Only a couple of these sources
might have a luminosity in the QSO range (Caccianiga private communication).
A similar interesting object with narrow optical emission lines, high 
X-ray luminosity and no evidence for obscuration in the X-ray spectrum, 
namely XLEO J0332-2744, was recently published  in \cite{bra05}.
However, differently from the three NEP objects, XLEO J0332-2744 shows 
a strong (EW = 0.5 - 1.9 keV)  Fe-line which is not present in our X-ray
spectra. 

\medskip\noindent
Further observations are needed to strengthen our
findings. We need a better coverage of the H$\alpha$ region, we need to
measure the optical extinction directly from the optical continuum shape,
and possibly the  central Black Hole mass that is expected
to be in the 10$^{10}$ M$_{\odot}$ range. Ultimately higher energy 
X-ray observations will say the final word on the presence of a 
Compton  thick absorber in these objects.

\begin{acknowledgements}

The work presented here is based on observations made with the 
Italian Telescopio Nazionale Galileo (TNG) operated on the island 
of La Palma by the Centro Galileo Galilei of the INAF (Istituto 
Nazionale di Astrofisica) at the Spanish Observatorio del Roque de 
los Muchachos of the Instituto de Astrofisica de Canarias
and on observations obtained with XMM-Newton, an ESA science mission with
instruments and contributions directly funded by ESA member states and NASA.
A special thank to Valentina Braito for performing the FLEX check.
We would like to thank our colleagues Luigi Foschini, Laura Maraschi, 
Nick Scoville and Paola Severgnini for many useful discussions. 
We thank an anonymous referee for a number of useful suggestions and
comments.
IMG thanks the hospitality  of the Institute for Astronomy of the 
University of Hawai$'$i where  part of this paper was written. She also 
notes that this work was done in spite of  the continued efforts by the 
Italian government to dismantle publicly-funded fundamental research. 
This work was partially supported  by NASA grant NNG05GA70G.

\end{acknowledgements}

\clearpage
\bibliographystyle{aa}
\bibliography{bib}


\end{document}